\documentclass[epj]{svjour}
\usepackage{graphics}
\input{psfig.sty}
\usepackage{graphicx}
\usepackage{dcolumn}

\newcommand{\ar}{\arrowvert}

\newcommand{\be}{\begin{equation}}
\newcommand{\ee}{\end{equation}}
\newcommand{\ba}{\begin{eqnarray}}
\newcommand{\ea}{\end{eqnarray}}

\begin{document}
\title{The ratio of viscosity to entropy density in a pion gas\\
satisfies the KSS holographic bound
}
\author{ Antonio Dobado, Felipe J. Llanes-Estrada 
\inst{1}
}                     
\institute{ Departamento de F\'{\i}sica Te\'orica I,  Universidad
Complutense, 28040 Madrid, Spain }
\date{Received: date / Revised version: date}
%

\abstract{ 
We evaluate the ratio of shear viscosity to entropy density in a pion 
gas employing the Uehling-Uehlenbeck equation and experimental 
phase-shifts parameterized by means of the $SU(2)$ Inverse Amplitude 
Method. We find that the ratio for this monocomponent gas stays well 
above the KSS $1/(4\pi)$ bound. We find similar results with other sets 
of phase shifts and conclude the bound is nowhere violated.
} 
\maketitle

Recently,  Kovtun, Son and Starinets \cite{Kovtun:2004de}
have conjectured a universal bound 
for the viscosity to entropy ratio of any one-component dilute gas based 
on gravity duality arguments and the Heisenberg uncertainty principle. 
In natural units the conjecture is that the quotient of the shear 
viscosity to the entropy density $ \eta/s $ is greater or equal
than $4\pi$, for any quantum field theory with one field alone.

A general proof does not exist but no counterexample is yet known 
either.
This has caused considerable stir in the heavy-ion collision community 
as RHIC experiments are providing us with a picture of strongly 
interacting matter \cite{Csernai:2006zz} that is close to a perfect 
fluid and features low viscosity. 

Prospects for an indirect measurement of viscosity in heavy-ion 
collisions through sufficiently precise hydrodynamic codes remain
appealing \cite{Teaney:2003kp}

Follow-up studies are under preparation by several groups from the 
hadron and from the quark-gluon phases in RHIC-theory to see the effect 
of the phase transition on the viscosity.

We have presented a comprehensive study of the shear viscosity 
of a hadronic gas at low and moderate temperature \cite{Dobado:2003wr}. 
Other transport coefficients are under active study, for 
example there is a recent paper featuring the electrical conductivity of 
the pion gas in chiral perturbation theory 
\cite{Fernandez-Fraile:2005ka} and a recalculation of the thermal 
conductivity is under preparation.
In this brief report we employ our published results to address the 
viscosity to entropy density ratio.

We employ the same notation and conventions as in our earlier 
publication \cite{Dobado:2003wr}. In particular we employ the $SU(2)$ 
Inverse Amplitude Method parametrization of the pion-pion scattering 
experimental phase shifts (as well as alternative parametrizations to 
check the sensitivity of the calculation presented).

To show a consistent ratio where both numerator and denominator, $\eta$ 
and $s=S/V$ are computed with the same approximation, we need to realize 
that the viscosity of the pion gas is defined hydrodynamically near 
equilibrium, that is, for infinitesimally small gradients of the gas 
velocity field. Therefore the same approximation can be taken in the 
calculation of the entropy, and this has to be evaluated at equilibrium 
employing the Bose-Einstein distribution function $f_0$.

Further, the  Uehling-Uehlenbeck (quantum Boltzmann) equation is 
decoupled from the BGKY hierarchy by assuming low density, that is, 
between two consecutive interactions, well separated by a large mean 
free path, particles decorrelate and behave as if the density of 
available states corresponded to a free gas.
Therefore it is also fair to employ the free gas approximation in the 
computation of the entropy density.

\begin{figure*}
\includegraphics[scale=0.50,angle=-90]{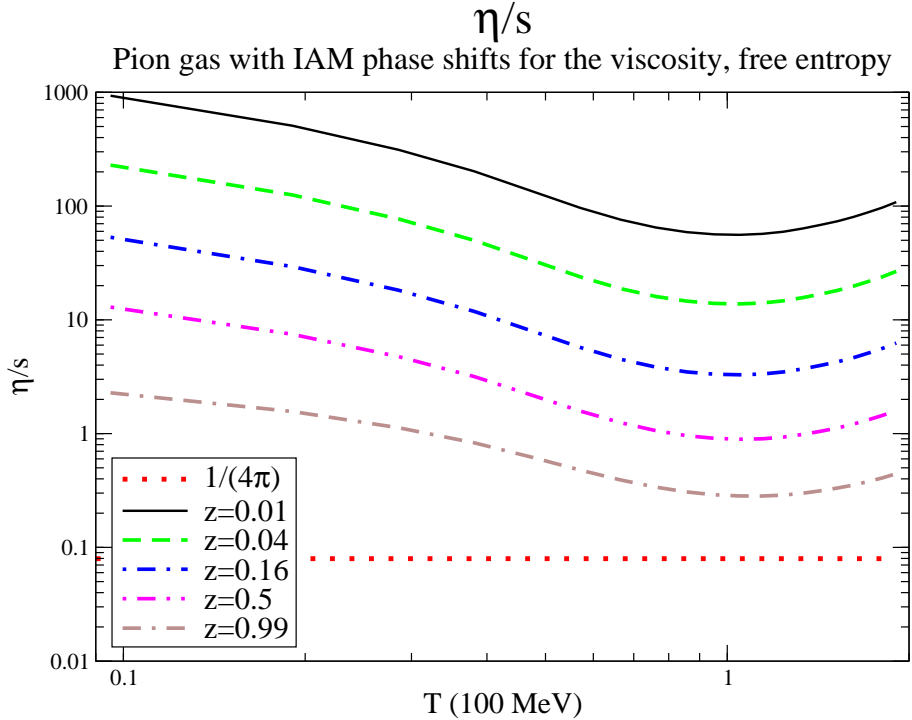}
\caption{ Viscosity over entropy density as a function of the
temperature for various chemical potentials (fugacities
$z=e^{(\mu_\pi-m_\pi)/T})$. We employ the IAM parametrization of the
pion-pion scattering phase shifts. As can be seen, the bound of Kovtun,
Son and Starinets is respected and no hint of a crossing or phase
transition appears from $\eta/s$ within this monocomponent pion gas }
\label{IAM}
\end{figure*}

Hence we are entitled to write 
\be
\log Z = - V g \int \frac{d{\bf{p}}}{(2\pi)^3} \log
\left[ 1-e^{-(E-\mu_\pi)/T}
\right]
\ee

Note that the chemical potential changes the density function $n(T)$
\begin{equation}
n\left(y,z \right) = \frac{g_{\pi}m_{\pi}^3}{4\pi^2}
\int_0^{\infty} dx \frac{x^{1/2}}{z^{-1} e^{y(\sqrt{1+x}-1)}
-1} 
\end{equation}
in terms of the fugacity $z=e^{(\mu_\pi-m_\pi)/T}$ and with 
$y=\frac{m_{\pi}}{T}$.

The total entropy follows from
\be
S=\frac{\partial (T\log Z)}{\partial T}
\ee
and the entropy density, after dividing by the volume and integrating 
once per parts, becomes
\be
s= \frac{g}{6\pi^2T^2} \int_0^\infty p^4dp 
\frac{e^{(E-\mu_\pi)/T}}{(e^{(E-\mu_\pi)/T}-1)^2} \ .
\ee

In figure \ref{IAM} we show the quotient of our previous computation of 
the viscosity divided by the entropy density for various temperatures 
and fugacities. It is clear that the holographic bound is at all times 
respected. Note that while the original paper 
\cite{Kovtun:2004de} had zero chemical potential, the recent $N=4$ 
super-Yang-Mills computation in \cite{Son:2006em} shows the bound is 
still valid for arbitrary $\mu$. This is indeed the case for the pion 
gas as we find. 

\begin{figure*}
\includegraphics[scale=0.50,angle=-90]{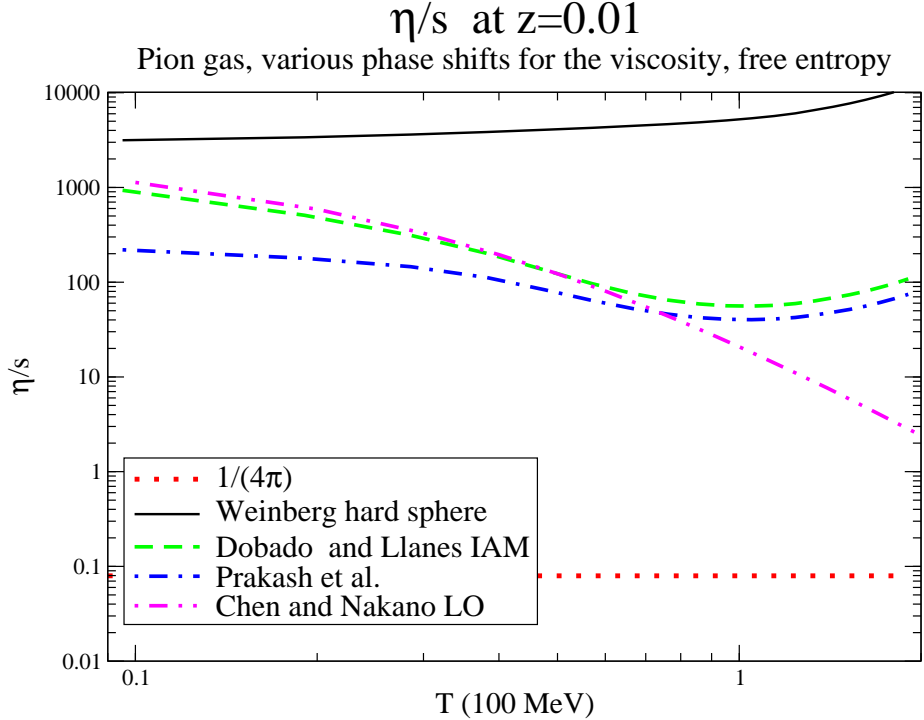}
\caption{
Viscosity over entropy density as a function of the
temperature for fixed fugacity $z=0.01$. We show various choices of 
interactions. The solid line at the top employs Weinberg's scattering 
lengths and the scattering amplitude is constant. The similar results 
of the IAM and the simple fit of Prakash et al. \cite{Prakash:1993bt}
for the scattering phase-shifts, naturally varying with momentum, mark 
what the ratio should be in the pion gas at moderate temperatures of 
order $50-100$ MeV. 
Finally, the LO result in Chiral Perturbation Theory of 
\cite{Chen:2006ig}, growing with the average momentum in a collision 
indefinitely as a polynomial, greatly diminishes the viscosity and 
therefore its quotient by the entropy density. } 
\label{z01} 
\end{figure*}

To check the dependence of this quotient to the interaction chosen, we 
plot in figure \ref{z01} the same $\eta/s$ but where the viscosity has 
been evaluated not only with the Inverse Amplitude Method (IAM) 
\cite{Dobado:1996ps}, but also with another reasonable parametrization 
of the scattering phase shifts \cite{Prakash:1993bt}, also with a
constant cross-section based on Weinberg's scattering lengths
\be
\sigma= \frac{23}{384} \frac{m_\pi^2}{\pi f_\pi^4}
\ee 
and finally with the full leading order in Chiral perturbation theory
\be
\ar T \ar^2 = \frac{1}{9f_\pi^4} (21 m_\pi^4 +9s^2 -24 m_\pi^2s
+3(t-u)^2) \ .
\ee

\begin{figure*}
\includegraphics[scale=0.50,angle=-90]{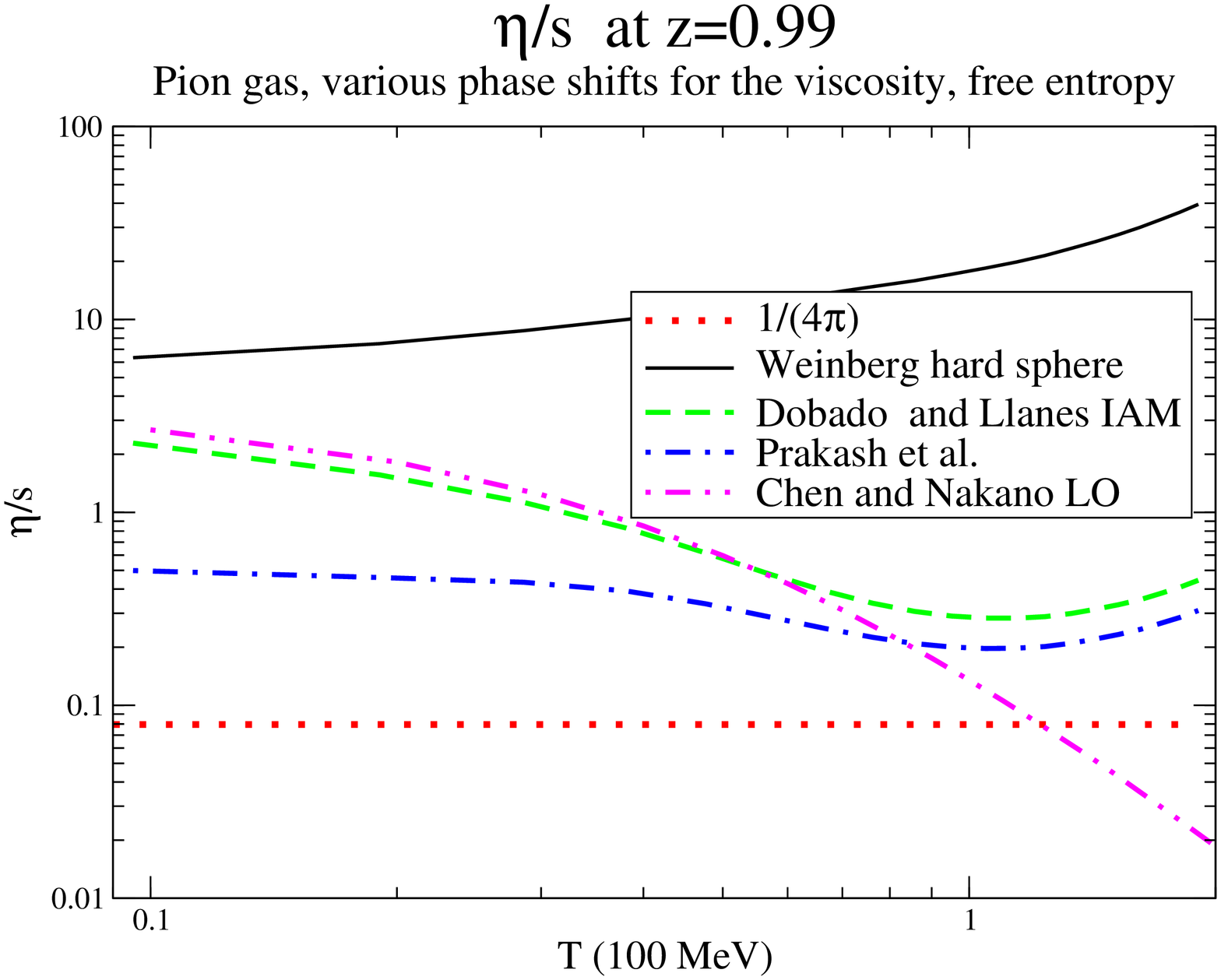}
\caption{
Same as in figure \ref{z01}.
The ratio computed in LO $\chi$PT \cite{Chen:2006ig}, growing with the 
average momentum in a collision
indefinitely as a polynomial, has an unphysical cross-over below the 
Kovtun-Son-Starinets bound that is not supported by realistic 
phase-shifts.}
\label{z99}
\end{figure*}

Employing the latter LO result, the authors of \cite{Chen:2006ig}
find a cross-over violating the $\frac{1}{4\pi}$ bound. 
We can discuss this by allowing the chemical potential to approach 
$m_\pi$ by choosing $z=0.99$, the results being plotted in figure 
\ref{z99}. There we see that this crossing is an artifact of the leading 
order Chiral Perturbation Theory and not to be expected from any 
realistic parametrization of the scattering phase shifts, cutting all 
speculation about a phase transition being visible from $\eta/s$ in this 
one-component gas. The reason is simple: unitarity at the $\rho$ pole in 
pion-pion scattering tames the amplitude growth that is unchecked in 
LO-$\chi$PT. This unphysical growth reflects in an ever-decreasing 
viscosity, sending $\eta/s$ to zero and violating any bounds.
The IAM correctly incorporates elastic unitarity and is free of this 
feature. The viscosity computed with this method is moreover comparable 
with published works \cite{Davesne:1995ms}.
Even at the highest fugacity, $\eta/s$ is a good factor of 2 
above the $\frac{1}{4\pi}$ bound.

\begin{figure*}
\includegraphics[scale=0.50,angle=-90]{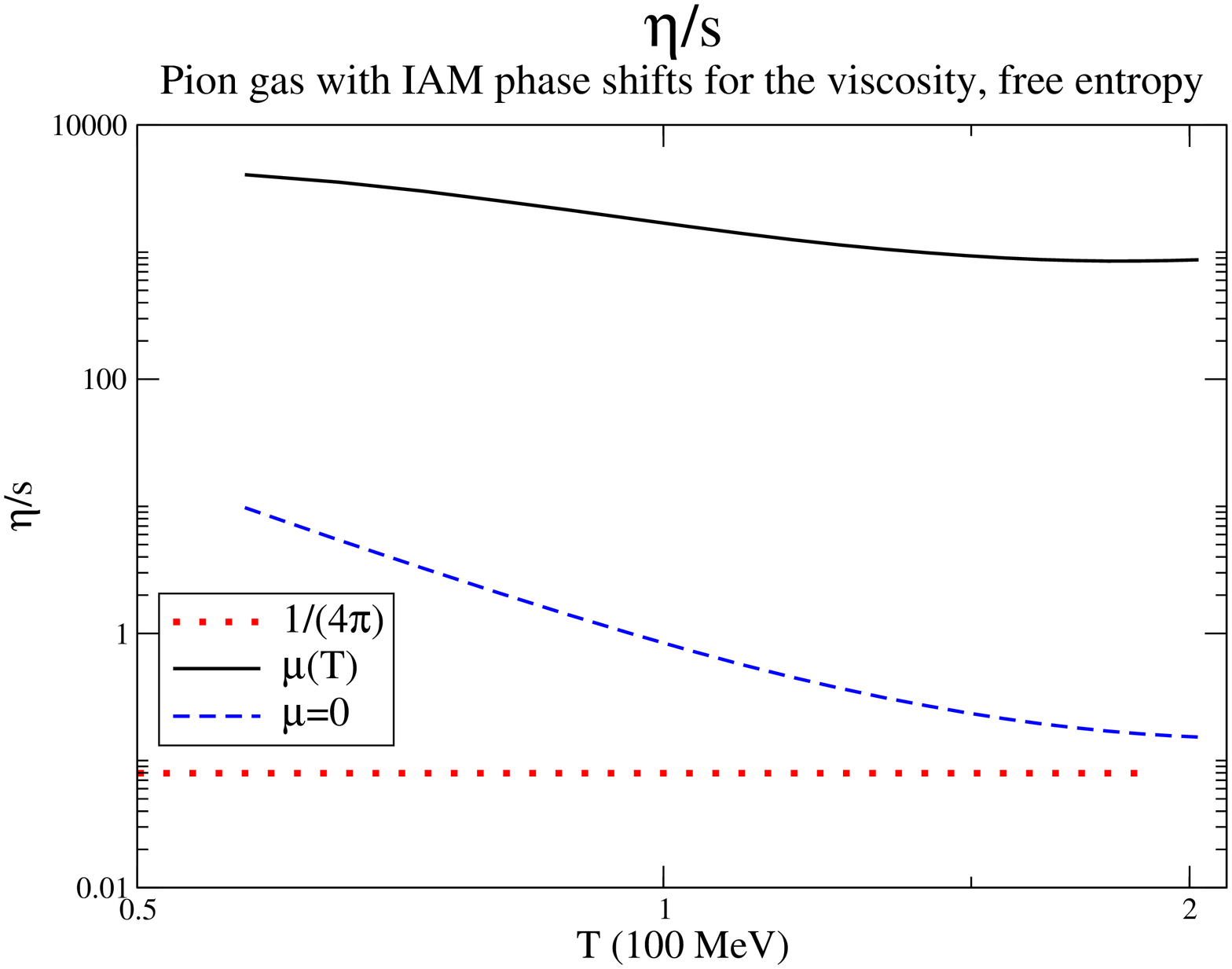}
\caption{Same as in figure \ref{z01} except we do not employ fixed
fugacity.
We plot the quotient $\eta/s$ as function of temperature for $\mu=0$ 
(for comparison) and for $\mu$ given in equation \ref{mudeT}, that is a 
rough guess of its evolution in a heavy-ion collision. Reading this 
graph from right to left we can see that the quotient increases with the 
expansion of the pion gas. The possibility of a minimum at a phase 
transition is not discarded and further work in this direction is 
necessary including other mesons, as approaching the critical 
temperature populates successively other flavors and spins (and 
eventually nucleons). } \label{mugraph} \end{figure*}

After ensuring that this bound is nowhere violated, one wonders whether
the observation \cite{Csernai:2006zz} that a minimum of $\eta/s$ 
might be a useful signature of a phase transition in Heavy Ion 
Collisions. To address this we need a model for the evolution of the 
pion chemical potential with the temperature. Then one can think of the 
temperature as a clock ticking the time after the phase transition in 
reverse.  
We can take the following model from ref. \cite{Liu:2006zy}, consistent 
with the thermal photon spectrum in such collisions, 
\be \label{mudeT}
\mu_\pi= \frac{2-T}{1.05}
\ee
(both $T$ and $\mu$ expressed in 100-MeV units).
The result is plotted in figure \ref{mugraph}. For comparison we also 
plot the fixed $\mu=0$ case.
As can be seen, the viscosity decreases smoothly and the possibility of 
a minimum at the phase transition is not ruled-out by the present 
analysis. To make this conclusion stronger one would need to populate 
further resonances in the hadron gas. This work will be undertaken in 
the near future.

\vspace{1cm }
{\it{We thank Jochen Wambach for encouraging us to carry on 
this interesting calculation, Juan M. Torres Rinc\'on for useful 
comments, and Angel Gomez Nicola, Daniel Fernandez 
Fraile and Dany Davesne for feedback on their own evaluation of 
transport coefficients in a pion gas. This work has been supported by 
grants FPA 2004-02602, 2005-02327, PR27/05-13955-BSCH (Spain).
}}

\end{document}